\documentclass[]{aa} 

\usepackage{graphicx, times, txfonts, float}

\newcommand{\less}{\raisebox{-1.1mm}{$\stackrel{<}{\sim}$}} 
 
\newcommand{\msol}{\mbox{M$_{\odot}$}} 
 
\newcommand{\msolyr}{{M$_{\odot}$}\,yr$^{-1}$} 
\newcommand{\mdot}{$\dot{M}$}
\newcommand{\lsol}{\mbox{L$_{\odot}$}} 
 
\newcommand{\kks}{K km s$^{-1}$} 
\newcommand{\ks}{km s$^{-1}$}

\begin{document}

\title{
The ALMA detection of CO rotational line emission in AGB stars in the Large Magellanic Cloud
}  
 
\author{ 
M.~A.~T.~Groenewegen\inst{1} 
\and
W.~H.~T.~Vlemmings\inst{2} 
\and
P.~Marigo\inst{3}
\and
G.~C.~Sloan\inst{4,5,6}
\and 
L.~Decin\inst{\ref{IvS}}
\and 
M.~W.~Feast\inst{\ref{UCT},\ref{SAAO}}
\and 
S.~R.~Goldman\inst{\ref{Keele}}
\and 
K.~Justtanont\inst{\ref{Chalmers}}
\and 
F.~Kerschbaum\inst{\ref{Vienna}}
\and 
M.~Matsuura\inst{\ref{Cardiff}}
\and 
I.~McDonald\inst{\ref{Jodrell}}
\and 
H.~Olofsson\inst{\ref{Chalmers}}
\and 
R.~Sahai\inst{\ref{JPL}}
\and 
J.~Th.~van~Loon\inst{\ref{Keele}}
\and 
P.~R.~Wood\inst{\ref{ANU}}
\and 
A.~A.~Zijlstra\inst{\ref{Jodrell}}
\and 
J.~Bernard-Salas\inst{\ref{OpenUniv}} 
\and 
M.~L.~Boyer\inst{\ref{GSFC},\ref{UMaryl}} 
\and 
L.~Guzman-Ramirez\inst{\ref{ESOc},\ref{Leiden}} 
\and 
O.~C.~Jones\inst{\ref{STSCI}}
\and 
E.~Lagadec\inst{\ref{OdCA}}
\and 
M.~Meixner\inst{\ref{STSCI}}
\and 
M.~G.~Rawlings\inst{\ref{EAO}}
\and 
S.~Srinivasan\inst{\ref{ASinica}}
}

\institute{ 
Koninklijke Sterrenwacht van Belgi\"e, Ringlaan 3, B-1180 Brussels, Belgium \\ \email{martin.groenewegen@oma.be}
\and
Department of Earth and Space Sciences, Chalmers University of Technology, Onsala Space Observatory, S-439 92 Onsala, Sweden\label{Chalmers}
\and
Department of Physics and Astronomy G.\ Galilei, University of Padova, Vicolo dell'Osservatorio 3, I-35122 Padova, Italy\label{Padova}
\and
  Cornell Center for Astrophysics \& Planetary Science, Cornell Univ., Ithaca, NY 14853-6801, USA\label{Cornell}
\and
  Department of Physics and Astronomy, University of North Carolina, Chapel Hill, NC 27599-3255, USA\label{UNorthC}
\and
  Space Telescope Science Institute, 3700 San Martin Dr., Baltimore, MD 21218, USA\label{STSCI}
\and 
Institute of Astronomy, Department of Physics and Astronomy, University of Leuven, Celestijnenlaan 200D, 3001 Leuven, Belgium\label{IvS}
\and 
  Astronomy Department, University of Cape Town, 7701, Rondebosch, South Africa\label{UCT} 
\and 
  South African Astronomical Observatory, P.O. Box 9, 7935, Observatory, South Africa\label{SAAO} 
\and 
Lennard-Jones Laboratories, Keele University, Staffordshire ST5 5BG, UK\label{Keele}
\and
Department of Astrophysics, University of Vienna, T\"urkenschanzstr. 17, A-1180 Vienna, Austria\label{Vienna}
\and 
School of Physics and Astronomy, Cardiff University, Queen's Buildings, The Parade, Cardiff, CF24 3AA, UK\label{Cardiff}
\and 
Jodrell Bank Centre for Astrophysics, School of Physics and Astronomy, University of Manchester, Oxford Road, Manchester M13 9PL, UK\label{Jodrell} 
\and 
Jet Propulsion Laboratory, MS\,183-900, California Institute of Technology, Pasadena, CA 91109, USA\label{JPL}
\and
Research School of Astronomy \& Astrophysics, The Australian National University, Canberra, ACT 2611, Australia\label{ANU}
\and 
Department of Physical Sciences, The Open University, MK7 6AA, Milton Keynes, UK\label{OpenUniv}
\and 
   Observational Cosmology Lab, Code 665, NASA Goddard Space Flight Center, Greenbelt, MD 20771 USA\label{GSFC}
\and 
   Department of Astronomy, University of Maryland, College Park, MD 20742 USA\label{UMaryl}
\and 
     European Southern Observatory, Alonso de C\'ordova 3107, Santiago, Chile\label{ESOc}
\and   
     Leiden Observatory, Leiden University, Niels Bohrweg 2, NL-2333 CA Leiden, The Netherlands\label{Leiden}
\and
Laboratoire Lagrange, Universit\'e C\^{o}te d'Azur, Observatoire de la C\^{o}te d'Azur, CNRS, Bd de l'Observatoire, CS 34229, F-06304, Nice Cedex 4, France\label{OdCA}
\and 
East Asian Observatory, 660 N. A'ohoku Place, University Park, Hilo, Hawaii 96720, USA\label{EAO}
\and 
Institute of Astronomy \& Astrophysics, Academia Sinica, 11F, Astronomy-Mathematics Building, No. 1, Roosevelt Rd., Sec 4, Taipei 10617, Taiwan (R.O.C.)\label{ASinica}
} 
 
\date{received: 2016, accepted: 2016} 
 
\offprints{Martin Groenewegen} 
 
\authorrunning{Groenewegen et al.} 
\titlerunning{CO rotational line-emission in AGB stars in the LMC} 
 
\abstract
%
   {Low- and intermediate-mass stars lose most of their stellar mass at the end of their lives on the asymptotic giant branch (AGB).
Determining gas and dust mass-loss rates (MLRs) is important in quantifying the contribution of evolved stars to the 
enrichment of the interstellar medium.}
   {Attempt to, for the first time, spectrally resolve CO thermal line emission in a small sample of AGB stars 
in the Large Magellanic Cloud.}
   {The Atacama Large Millimeter Array was used to observe 2 OH/IR stars and 4 carbon stars in the LMC in the CO J= 2-1 line.}
   {We present the first measurement of expansion velocities in extragalactic carbon stars. 
All four C-stars are detected and wind expansion velocities and stellar velocities are directly measured. 
Mass-loss rates are derived from modelling the spectral energy distribution and {\it Spitzer}/IRS spectrum with the DUSTY code.
Gas-to-dust ratios are derived that make the predicted velocities agree with the observed ones.
The expansion velocities and MLRs are compared to a Galactic sample of well-studied relatively low MLRs stars supplemented 
with ``extreme'' C-stars that have properties more similar to the LMC targets.
Gas MLRs derived from a simple formula are significantly smaller than derived from the dust modelling, indicating 
an order of magnitude underestimate of the estimated CO abundance, time-variable mass loss, or that the CO intensities 
in LMC stars are lower than predicted by the formula derived for Galactic objects. This could be related to 
a stronger interstellar radiation field in the LMC.
 }
   {Although the LMC sample is small and the comparison to Galactic stars is non-trivial because of uncertainties 
in their distances (hence luminosities) it appears that for C stars the wind expansion velocities in the LMC are lower than in the 
solar neighbourhood, while the MLRs appear similar. 
This is in agreement with dynamical dust-driven wind models. }


\maketitle

\section{Introduction} 

Low- and intermediate-mass stars (LIMS) have initial masses of $\sim$0.8--8M$_{\odot}$, depending somewhat on metallicity.
They end their lives with an intense mass-loss episode on the asymptotic giant branch (AGB). 
In a classical picture (Wood 1979), stellar pulsation and dust formation drive a slow, cool wind. This wind 
from AGB stars is one of the main sources that enrich the interstellar medium (ISM) with gas and dust.

To quantify the level of enrichment and compare it to other sources, like supernovae or dust growth in the ISM, one has
to determine the mass-loss rate (MLR) in gas and dust of AGB stars. Typically, the dust MLR is determined 
by modelling the spectral energy distribution (SED) and is especially sensitive to the 
infrared photometry. The gas MLR is determined by modelling 
the rotational-vibrational transitions of carbon monoxide (CO).
The dust MLR  is directly proportional to the (dust) expansion velocity of the wind, which is a priori unknown, 
and can not be determined from the SED fitting.
Expansion velocities are known for hundreds of AGB stars in the Galaxy, through the CO thermal line emission 
(e.g. Kerschbaum \& Olofsson 1999, Olofsson et al.\ 2002 for M-stars, Groenewegen et al.\ 2002 for C-stars) or
OH maser line for O-rich sources (see the database by Engels \& Brunzel 2015).

For stars that are beyond a few kpc, determining expansion velocities becomes more difficult especially in the CO line.
In part this is a sensitivity issue of the receivers and telescopes; in part the AGB population that one traces 
(outside the Galactic disk for example) may be constituted of stars of lower mass that may have an inherently lower MLR and 
thus fainter line emission.

The first indication that the wind expansion velocities of AGB stars depend on environment (metallicity) came with the detection
of C-stars in the Galactic Halo. Groenewegen et al. (1997) detected CO (2-1) emission in the source IRAS 12560+1656, while 
Lagadec et al. (2010) detected CO J= 3-2 emission in that source and five others. 
The expansion velocities were lower than that of C-stars in the Galactic disk, with the expansion velocity 
in IRAS 12560 as low as $\sim 3$ \ks.

Hydrodynamical wind models for C stars by Wachter et al.\ (2008) predict that expansion velocities in the LMC
are indeed about 2.2 $\pm$ 0.2 times lower than for solar metallicities, while dust-to-gas ratios are 
slightly larger (by a factor of $\sim$1.3), and mass-loss rates are relatively unchanged.

Since maser emission is stronger than thermal emission, attempts have been made to detect OH emission in the 
Small and Large Magellanic Clouds (MCs). Wood et al. (1992) reported the detection in 6 
AGB and RSG stars observed in the LMC 
and no detection in 
the SMC.
Marshall et al. (2004) added a few detections in the LMC. The current state-of-the-art is 
presented by Goldman et al. (2016), adding new detections and re-analysing previous data, resulting in accurate expansion velocities
for 13 OH/IR stars in the LMC. They suggest that the expansion velocity is proportional to metallicity and luminosity, $L$, as $L^{0.4}$.

Recently, Matsuura et al. (2016) detected CO line emission in the bright 
LMC RSGs WOH G64 (one line) and IRAS 05280$-$6910 (J= 6-5 to 15-14).
The data were obtained using the {\it Herschel} PACS and SPIRE instruments. The spectral resolution is insufficient to 
resolve the lines, but, taking the wind expansion velocity from an OH measurement, they modelled the CO line emission 
in IRAS 05280 and derived a MLR.

It would be extremely interesting to extend this type of analysis to carbon-rich stars and to larger samples so that a
meaningful comparison to Galactic objects could be made. The collecting area and small beam of the Atacama Large Millimeter Array (ALMA) is 
required to detect spectrally resolved CO lines in the MCs.

This paper describes the first results of such a programme.
The target list is presented in the next section.
The ALMA observations and results are presented in Section~3.
These results are discussed in Section~4 where they are compared to predictions of radiatively driven wind theory, and 
the results for the LMC are compared to a Galactic sample. The evolutionary status of the stars is examined with the aid of AGB stellar models. 
Conclusions are presented in Section~5.

\section{Targets} 

Four carbon (C) stars and two oxygen-rich AGB stars were selected from the sample of about 
225 C and 170 M stars studied by Groenewegen \& Sloan (in prep.; hereafter GS16).  This sample constitutes essentially all AGB stars and 
red supergiants (RSG) in the SMC and LMC observed with the Infrared Spectrograph (IRS) on {\it Spitzer}.
GS16 constructed the SED and fitted the SED and IRS spectrum with the 
radiative transfer (RT) code ``More of DUSTY'' (MoD, Groenewegen 2012). 
The fitting procedure estimates luminosity and (dust) mass-loss rate.
In addition GS16 determined the pulsation periods for many of the stars based on a (re-)analysis of available multi-epoch photometry. 
This included the OGLE, MACHO, EROS photometric survey data, $K-$band data available in the literature for the redder sources, 
combined with $K-$band epoch photometry from the VMC survey (Cioni et al. 2011), and 
data in the mid-IR from the SAGE-Var programme (Riebel et al. 2015, mostly the 4.5 $\mu$m data), and the W2 filter data from the 
AllWISE multi-epoch photometry (Wright et al. 2010) and NEO-WISE mission (Mainzer et al. 2011, see also Sloan et al. 2016).

Simple formulae\footnote{and supported by detailed RT calculations carried out in the preparation of the ALMA proposals.} 
like Olofsson (2008) indicate that the expected flux in the low-level transitions of CO is proportional 
to $\sim$ \mdot$^{1.2} f_{\rm CO}^{0.7} $, where $f_{\rm CO}$ is the abundance of CO relative to H$_2$.
The best estimate for the MLR in making the line-flux predictions comes from modelling the dust (GS16), but 
then an uncertain dust-to-gas ratio (DTG) must be assumed. Given all these uncertainties and the sensitivity of ALMA it was 
evident that only the sources with the largest MLR estimates could potentially be detected.

Table~\ref{Tab-Targets} lists the final sample.
As an indication of their redness, the 2 OH/IR stars have the 4th and 6th largest estimated dust optical depth of the 170 M stars 
studied in GS16, while the 4 C-stars are ranked between 2nd and 17th among 225 C-stars. 
The dust optical depth, luminosity and pulsation properties (period, (semi-)amplitude and 
in which filter) are taken from GS16 (see also Sloan et al. 2016). 
All objects are large-amplitude variables (LPVs).

\begin{figure*}[ht]
\centering

\includegraphics[width=0.85\hsize]{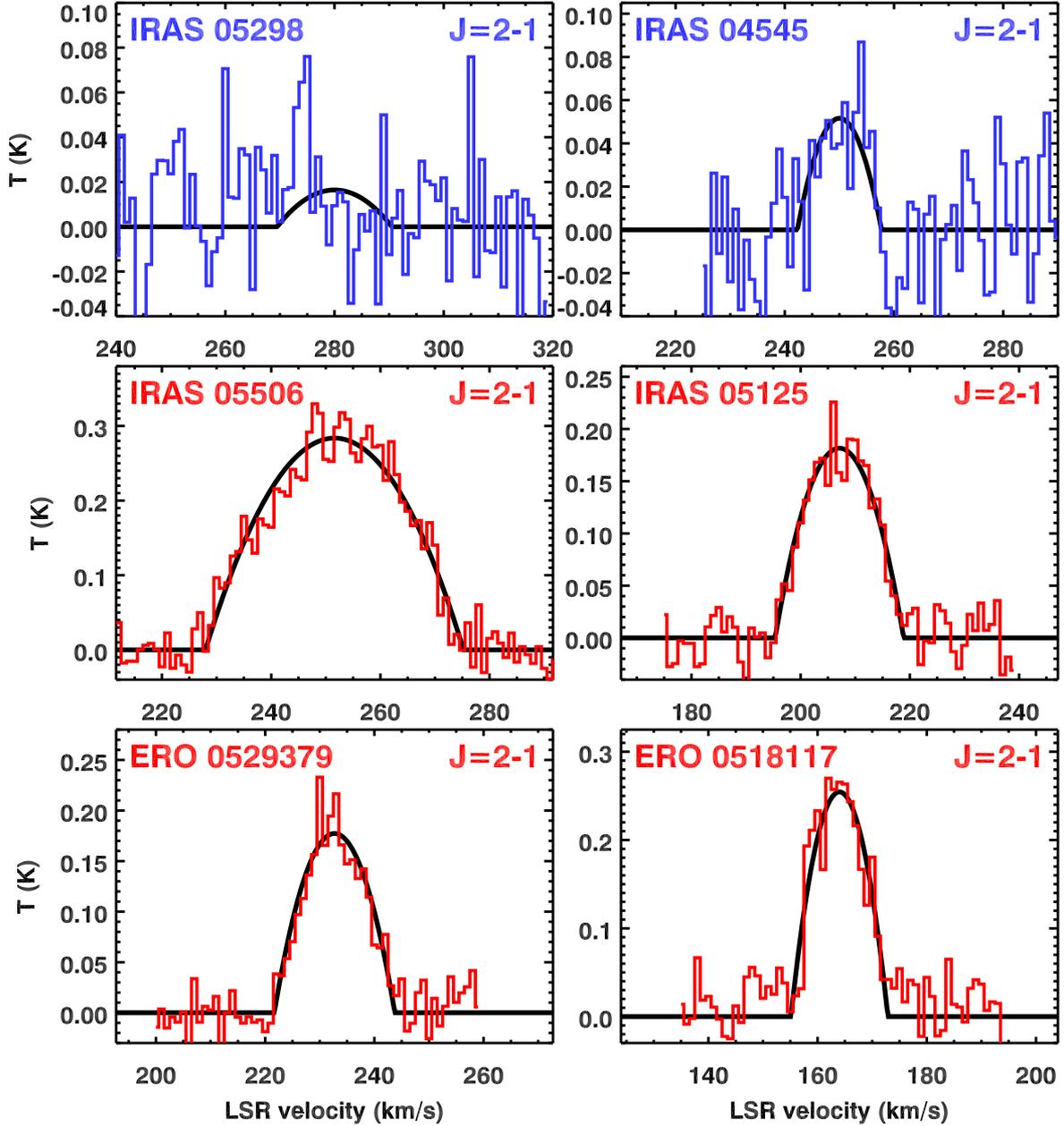}

\caption[]{ 
CO (2-1) profiles of the 2 OH/IR (top, in blue) and 4 carbon stars.
The solid line is a fit to the data using Equation~1.
The fit to IRAS 05298 is not significant, see Table~\ref{Tab-Fit}.
} 
\label{Fig-Prof} 
\end{figure*}

\begin{table*}

 \caption{LMC targets and some properties.}
\centering
  \begin{tabular}{lcccccccccccccc}
  \hline\hline
Identifier  & IRAS name &    RA     &   DEC     &  $\tau_{\rm d}$ &     Lum.   & Period  & (Semi-)Amplitude (Filter) \\ 
            &           &   (deg)   &  (deg)    &                &    (\lsol) & (days)  &   (mag)          \\ 
\hline
\multicolumn{8}{c}{OH/IR stars} \\
\hline
IRAS 05298  & 05298$-$6957 & 82.35260  & $-$69.92047 & 25.8  & 37~700 & 1265 & 0.98 ($K$) \\   
IRAS 04545  & 04545$-$7000 & 73.54188  & $-$69.93283 & 29.5  & 24~900 & 1254 & 0.85 ($K$) \\   
\hline
\multicolumn{8}{c}{carbon-rich stars} \\
\hline
IRAS 05506  & 05506$-$7053 & 87.48550  & $-$70.88658 & 24.5 & 17~800 & 1026 & 0.67 ($K$) \\  
IRAS 05125  & 05125$-$7035 & 78.00321  & $-$70.54000 & 43.8 & 15~500 & 1115 & 0.61 ($K$) \\ 
ERO 0529379 & 05305$-$7251 & 82.40786  & $-$72.83136 & 172. & \phantom{1}5~400 & 1076 & 0.49 (WISE W2) \\ 
ERO 0518117 & 05187$-$7033 & 79.54878  & $-$70.50750 & 79.2 & \phantom{1}9~300 & 1107 & 0.45 (IRAC Ch2) \\ 
\hline
\end{tabular}

\tablefoot{Dust optical depth, $\tau_{\rm d}$, at 0.5 $\mu$m, luminosity and variability information (from GS16). \\
}

\label{Tab-Targets}
\end{table*}

\section{Observations and analysis} 

The sources were observed with ALMA at Band 6 on 2015 January 18, 19, and 
April 9 using 34, 41, and 35 12m antennas, respectively. 
The observations have four spectral windows; one window
with a width of $937.5$~MHz and 3840 channels to cover the CO(2-1)
transition, and three $2$~GHz windows with 128 channels each for the
continuum. The line spectral window was centered on 230.303~GHz, and
the continuum windows were centered on 212.791, 214.789, and 227.782~GHz. 
Because of Hanning smoothing, the effective velocity
resolution in the line spectral window was $\sim0.64$~km~s$^{-1}$. The
observations covered a baseline range of $\sim15-350$~m providing a
maximum recoverable scale of $\sim10.5\arcsec$. The total observing time was
3.9~hrs of which 2.1~hrs were spread equally over the six sources.

We calibrated the data using the CASA 4.2.2 pipeline.
Manual calibration with CASA 4.5.2 yielded consistent results. 
Ganymede and Callisto were used as flux calibrators, 
and the quasars J1107$-$4449, J0750+1231, J0519$-$4546, and J0538$-$4405 were used for 
bandpass calibration depending on the date of observation. 
The quasar J0635$-$7516 was used for complex-gain calibration at all observing dates.

After calibration, continuum images were produced using natural weighting. 
Where needed, continuum subtraction was performed and the
CO(2-1) line was imaged averaging to $1$~km~s$^{-1}$ spectral
resolution. Both for continuum and the line images, the typical beam
size is $1.7\arcsec\times1.2\arcsec$, at a position angle of $\sim$$80^\circ$. The
typical rms noise in the continuum images is $\sim$45$\mu$Jy~beam$^{-1}$
and that in the line maps ranges from $3.1-4.1$~mJy~beam$^{-1}$.

The maximum baseline (and thus the size of the beam) was chosen to match the expected CO photodissociation radius 
based on the formula of Mamon et al. (1988) so that the emission region remains unresolved.

\medskip

Line profiles were extracted from an area typically twice the beam-size and converted to a temperature scale within CASA.
Figure~\ref{Fig-Prof} shows the observed line profiles.  All 4 C stars are detected, and one of the two OH/IR stars.
The line profiles were fitted with a ``Shell'' profile as defined in the CLASS/GILDAS software package
\footnote{https://www.iram.fr/IRAMFR/GILDAS/doc/html/class-html/node38.html}:

\begin{equation}
   P(V) = \frac{A}{\Delta V \; (1 + H /3)} \; \left(1 + 4 H \; \left(\frac{V - V_{\star}}{\Delta V}\right)^2\right),
\end{equation}
where $V_{\star}$ is the stellar velocity (in \ks, throughout the paper the LSR frame is used), $A$ is the integrated intensity (in \kks), 
$\Delta V$ the full-width at zero intensity (in \ks, and the expansion velocity v$_{\rm exp}$ is taken as half that value), 
and $H$ the horn-to-center parameter. This parameter describes the shape of the profile, and 
is $-1$ for a parabolic profile, $0$ for a flat-topped one, and $>0$  for a double-peaked profile.
$H$ was fixed to $-1$ in the fitting. For IRAS 05506, the star with the best determined profile, we also fitted a profile 
allowing this parameter to vary as well and found $H= -1.02 \pm 0.02$.
A parabolic profile indicates optically thick, unresolved emission.
Table~\ref{Tab-Fit} lists the results. 
For the 2 OH/IR stars the stellar and expansion velocity were fixed to the values determined from 
OH observations (Goldman et al. 2016) because of the poor signal-to-noise ratio in the spectra.
The results for IRAS 05298 are not significant ($T_{\rm peak} = 0.016 \pm 0.009$K), and 3$\sigma$ upper limits are reported.

\begin{table*}

 \caption{Results of the CO J= 2-1 observations. }
\centering
  \begin{tabular}{lcccccc}
  \hline\hline
Identifier  &    v$_{\star}$ & v$_{\rm exp}$ &  $A$   & $T_{\rm peak}$ & rms  & gas \mdot\tablefootmark{b} \\
            &   (\ks)      &   (\ks)    &  (\kks)  &    (K)       & (K)  &     (\msolyr)    \\
\hline
\multicolumn{6}{c}{OH/IR stars} \\
\hline
IRAS 05298  &  280\tablefootmark{a} & 10.5\tablefootmark{a} & $< 0.36$ & $< 0.027$ & 0.028 & $<$2.8 $\times 10^{-6}$ \\ 
IRAS 04545  &  250\tablefootmark{a} & \phantom{1}7.7\tablefootmark{a} & 0.53 $\pm$ 0.09 & 0.053 $\pm$ 0.009 & 0.026 & 3.3 $\times 10^{-6}$ \\ 
\hline
\multicolumn{6}{c}{carbon-rich stars} \\
\hline
IRAS 05506  &  251.6 $\pm$ 0.3 & 23.63 $\pm$ 0.42 & 8.94 $\pm$ 0.19 & 0.284 $\pm$ 0.006 & 0.027 & 3.5 $\times 10^{-5}$ \\ 
IRAS 05125  &  207.1 $\pm$ 0.3 & 11.77 $\pm$ 0.15 & 2.85 $\pm$ 0.12 & 0.183 $\pm$ 0.007 & 0.024 & 1.0 $\times 10^{-5}$\\ 
ERO 0529379 &  232.7 $\pm$ 0.3 & 11.04 $\pm$ 0.41 & 2.61 $\pm$ 0.09 & 0.178 $\pm$ 0.007 & 0.019 & 8.9 $\times 10^{-6}$\\ 
ERO 0518117 &  164.0 $\pm$ 0.3 & \phantom{1}8.87 $\pm$ 0.52 & 3.01 $\pm$ 0.15 & 0.257 $\pm$ 0.013 & 0.033 & 9.0 $\times 10^{-6}$ \\ 
\hline
\end{tabular}
\tablefoot{
\tablefoottext{a}{Stellar velocity and expansion velocity for the OH/IR stars from Goldman et al. (2016), and fixed in the fitting.}\\
\tablefoottext{b}{Gas MLR based on Eq.~\ref{Eq-CO}, see text for assumed parameters.}
}
\label{Tab-Fit}
\end{table*}

\section{Discussion} 

In the remainder of this paper we will focus mainly on the outflow velocities from the carbon stars. 
For the OH/IR stars the expansion velocities are 
better determined from the OH profiles (Goldman et al. 2016), and for all 6 sources a detailed analysis of the CO line strength 
in terms of the (gas) MLR is deferred until the J= 3-2 data are in hand.
Goldman et al. discuss the 13 known OH/IR stars in the LMC (there are still none known in the SMC) and 
compare the expansion velocities to Galactic counterparts.

\subsection{Predictions by dust-driven wind theory} 

The DUSTY code can be run in ``density type = 3'' mode, i.e. the mode where the hydrodynamical equations 
of dust and gas are solved and the gas expansion velocity and gas MLR are predicted (Ivezi\'c \& Elitzur 1995, 2010).
The velocities and MLR given by DUSTY scale as $(L/10^{4})^{0.25} \left((r_{\rm gd}/200)(\rho_{\rm d}/3)\right)^{-0.5}$ and 
$(L/10^{4})^{0.75} \left((r_{\rm gd}/200)(\rho_{\rm d}/3)\right)^{0.5}$, respectively, 
where $L$ is the luminosity in solar units, $r_{\rm gd}$ the gas-to-dust ratio, and $\rho_{\rm d}$ the specific density of 
the dust grains in g cm$^{-3}$.

MoD (Groenewegen 2012) was modified to fit photometric and spectroscopic data when running DUSTY in ``density type = 3'' mode.
The fitted parameters are luminosity, dust optical depth, and dust temperature at the inner radius. 
This is one parameter less than the  version used by GS16 where the slope of the density 
law ($\sim r^{-p}$) could  also be fitted.
In ``density type = 3'' mode the density varies as $\sim r^{-2}$ once the terminal velocity is reached.

Table~\ref{TabDUSTY} presents the results. Appendix~\ref{AppDUSTY} shows the fitted models.
The luminosities agree well with the values determined by GS16 (as repeated in column~6 in Tab.\ref{Tab-Targets}).
The next column lists the MLR and expansion velocity for a gas-to-dust ratio of 200, scaled to the actual luminosity and 
grain density used in the fitting.
One can then compare this predicted expansion velocity to the observed value and determine the gas-to-dust ratio that makes them equal 
(reported in col.~5) and then scale the MLR using this value (reported in col.~6).
The results do not depend too strongly on the luminosity, $r_{\rm gd} \sim \sqrt{L}$ and $\dot{M}_{\rm scaled} \sim L^{0.25}$.
The statistical fitting error in the luminosity is about 10\%, to which one should add the uncertainty in the distance ($L \sim d^2$).
The DUSTY manual states that the MLR calculated in the ``density type = 3'' mode has an inherent uncertainty of $\sim$30\%.
The implications are further discussed below.

\begin{table}
\setlength{\tabcolsep}{1.7mm}

 \caption{DUSTY modelling of the LMC carbon stars.}
\centering
  \begin{tabular}{lrrrccccc}
  \hline\hline
Identifier  & Lum.    & MLR$_{\rm DUSTY}$    &  v$_{\rm exp, DUSTY}$  & $r_{\rm gd}$ &  MLR$_{\rm scaled}$  \\
            & (\lsol) & (\msolyr)         &      (\ks)          &            &  (\msolyr)    \\
\hline
IRAS 05506  &  17~700  & 2.0  $\times 10^{-5}$ & 19.3 & 133 &  1.6 $\times 10^{-5}$ \\ 
IRAS 05125  &  15~500  & 2.4  $\times 10^{-5}$ & 19.4 & 541 &  4.0 $\times 10^{-5}$ \\ 
ERO 0529379 &   5~600  & 2.9  $\times 10^{-5}$ & 17.5 & 504 &  4.5 $\times 10^{-5}$ \\ 
ERO 0518117 &   9~500  & 4.3  $\times 10^{-5}$ &  7.5 & 142 &  3.6 $\times 10^{-5}$ \\ 

\hline
\end{tabular}
\tablefoot{
Columns~3 and 4 give the gas MLR and gas expansion velocity for a gas-to-dust ratio of 200. 
Column~5 gives the gas-to-dust ratio that makes the observed expansion velocity equal to the observed one, and the last column lists 
the gas MLR for that gas-to-dust ratio.
}
\label{TabDUSTY}
\end{table}

\subsection{A Galactic comparison sample}

Because the dust-driven wind theory predicts that the expansion velocity depends on the gas-to-dust ratio, it is of interest to 
compare expansion velocities, luminosities and MLRs of the LMC targets to Galactic C-stars.

Expansion velocities exist for hundreds of Galactic AGB stars, but reasonable estimates of distances (and thus luminosities), 
and, especially, MLR estimates derived from detailed modelling are much rarer.
The Galactic sample of ``ordinary'' C-stars is taken from Ramstedt \& Olofsson (2014) and Danilovich et al. (2015), who 
both derive MLR estimates from multi-transitional CO data and detailed RT modelling.

Both samples contain 19 C-stars, of which two are in common, V384 Per and R Lep. 
Distances are taken from the literature and based on a variety of methods that include 
revised {\it Hipparcos} parallax data, or the period-luminosity ($PL$) relation 
from Groenewegen \& Whitelock (1996) for Miras\footnote{Only two of these 36 (and none of the 7 ``extreme'' C-stars discussed below) 
have a parallax in the Gaia Data Release 1 (Lindegren et al. 2016), with values that are consistent within 
the errors with the adopted distances in the literature.}.
The model is fitted to ground-based CO J= 1-0, 2-1, 3-2, and 6-5 data, while Danilovich et al. (2015) 
additionally use {\it Herschel}/HIFI J= 5-4 and 9-8 data (and more transitions for selected sources).
Both studies model the SEDs with DUSTY to provide the dust optical depth and dust temperature profile that enters in the 
calculations of the CO line transfer, and both studies use the Monte Carlo RT code developed by Sch\"oier \& Olofsson (2001).

These two papers provide a sizeable sample of uniformly modelled AGB stars with probably the most accurately 
determined MLRs currently available in the literature, but the samples contain only one source as red as the LMC targets.

Therefore we also considered the $\sim$30 sources classified as ``extreme carbon stars'' (Volk et al. 1992, ``group V'' sources from Groenewegen  et al. 1992).
These sources have been identified based on IRAS photometry and the shape of the IRAS LRS spectrum. Later, Speck et al. (2009) studied
10 of them (one new) using superior {\it ISO}/SWS spectra. Many of these sources have the silicon carbide feature in absorption, 
as seen in ERO 0529379.

Distances are a challenge for Galactic sources. 
Therefore, we have selected only the seven extreme carbon stars with known pulsation periods.

Table~\ref{TabExtr} lists the sample of extreme carbon stars. AFGL 3068 is the only star in common with the comparison sample from Ramstedt \& Olofsson (2014).
All stars are LPVs. The longest period is similar to the three LMC targets, and the periods range between $\sim$700 and 1060 days.
The CO-based expansion velocities in column~5 are from Groenewegen et al. (2002).
Luminosities have been determined using the $PL$ relation of Groenewegen \& Whitelock (1996), as for many stars in the local C-star sample.

The SED and mid-IR spectra ({\it ISO}/SWS or {\it IRAS}/LRS) are fitted using the modified MoD code using the ``density type = 3'' mode, 
as for the LMC targets. For the assumed luminosity, the best-fitting distance (col.~6) is determined.
The remaining 4 columns have identical meanings as for the LMC targets: the expansion velocity and MLR for a gas-to-dust ratio of 200, 
the  gas-to-dust ratio required to obtain the observed velocity, and the MLR for that gas-to-dust ratio.
The models fitted to SEDs and spectra are shown in Appendix~\ref{AppDUSTYExt}.

\begin{table*}
\setlength{\tabcolsep}{1.5mm}

 \caption{Galactic extreme carbon stars. }
\centering
  \begin{tabular}{llrccccccrcc}
  \hline\hline
Identifier  & Other name &  Period  & Amp. (Fil.) & Ref. & v$_{\rm exp}$ &   Lum.  &   d   & MLR$_{\rm DUSTY}$   &  v$_{\rm exp, DUSTY}$ & $r_{\rm gd}$ &  MLR$_{\rm scaled}$     \\
            &            &  (days)  &   (mag)     &      &   (\ks)     & (\lsol) & (kpc) & (10$^{-5}$\msolyr) &       (\ks)       &            &  (10$^{-5}$\msolyr)     \\
\hline
AFGL 190        & 01144+6658 & 1060 & 1.2 ($L$) & 1 & 18.0 & 16~400 & 3.30 & 5.7 & 10.5 &  67 & 3.3 \\ 
AFGL 341        & 02293+5748 &  815 & 0.9 ($L$) & 1 & 14.2 & 12~500 & 2.81 & 2.6 & 14.2 & 199 & 2.6 \\ 
IRAS 03448+4432 & AFGL 5102  &  729 & 0.7 ($K$) & 2 & 13.3 & 11~100 & 2.31 & 1.4 & 20.6 & 477 & 2.2 \\ 
AFGL 865        & 06012+0726 &  696 & 0.9 ($K$) & 3 & 16.6 & 10~600 & 1.72 & 1.5 & 17.9 & 234 & 1.6 \\ 
IRAS 08074-3615 & V688 Pup   &  832 & 0.9 ($K$) & 4 & 21.7 & 12~700 & 2.95 & 2.0 & 19.2 & 156 & 1.7 \\ 
AFGL 2494       & 19594+4047 &  783 & 1.0 ($L$) & 5 & 20.5 & 12~000 & 1.50 & 1.4 & 20.0 & 191 & 1.3 \\ 
AFGL 3068       & 23166+1655 &  696 & 0.9 ($K$) & 3 & 15.1 & 10~600 & 1.05 & 3.9 & 13.9 & 168 & 3.6 \\ 

\hline
\end{tabular}
\tablefoot{
References for the pulsation periods and semi-amplitudes are: 
(1) Groenewegen et al. (1998), based on $L$-band data from R.R. Joyce (priv. comm. 1996), 
(2) Kerschbaum et al. (2006),
(3) Le Bertre (1992),
(4) Whitelock et al. (2006),
(5) Jones et al. (1990).  \\ 
The output from MoD for the gas MLR and expansion velocity scale with $(r_{\rm gd}/200)^{p}$ with power $p= 0.5$ and $-0.5$, respectively.
}

\label{TabExtr}
\end{table*}

\subsection{A preliminary synthesis}

\begin{figure}
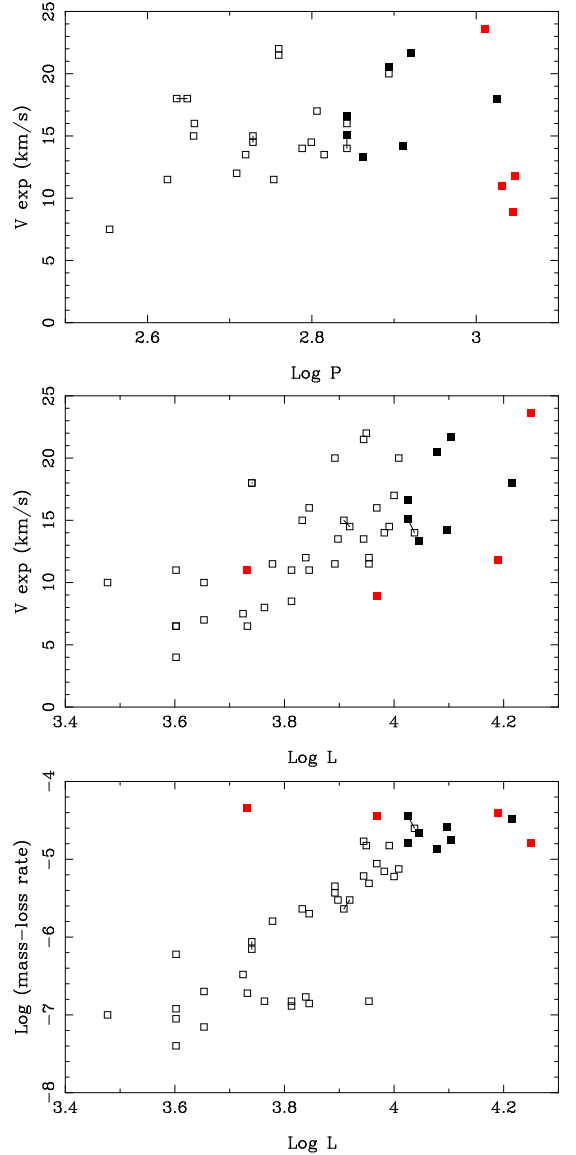

\centering

\includegraphics[width=0.8\hsize]{TD_V_Per_paper.ps}
\includegraphics[width=0.8\hsize]{TD_V_Lum_paper.ps}
\includegraphics[width=0.8\hsize]{TD_Mdot_Lum_paper.ps}

\caption[]{ 
Plotted are the 4 C-stars (filled red squares) in the LMC, and the Galactic C-stars from  Ramstedt \& Olofsson (2014) and 
Danilovich et al. (2015) 
(open squares), and the Galactic extreme C-stars (filled black squares).
The stars in overlap in the local sample are connected (R Lep, $\log P \sim 2.64, \log L \sim 3.75$, and 
V384 Per,  $\log P \sim 2.72, \log L \sim 3.92$), as well as the two independent estimates for 
AFGL 3068 ($\log P \sim 2.84, \log L \sim 4.03$).
} 

\label{Fig-Comp} 
\end{figure}

Figure~\ref{Fig-Comp} compares the LMC targets to the Galactic sample of ``standard'' and ``extreme'' C-stars.
The panels show the expansion velocity against $\log P$ (for the local sample, only Miras were selected in this case, 
as the Galactic extreme C-stars and the LMC targets are LPVs), expansion velocity against $\log L$, 
and MLR versus $\log L$.
The luminosities for the LMC stars are derived from the SED fitting (and an assumed distance), 
while for the ``extreme'' C stars they are derived from the $PL$ relation by Groenewegen \& Whitelock (1996). 
For 3 of the 4 LMC stars the luminosity derived from the modelling is consistent with this $PL$-relation.
For periods between 1026 and 1150 days the $PL$ relation predicts luminosities between 15~800 and 17~300 \lsol, in good
agreement with the luminosities of IRAS 05506 and 05125. The scatter in the observed $PL$ relation is 0.26 mag, 
and thus the luminosity of ERO 0518117 is consistent with the $PL$ relation at the 2$\sigma$ level. ERO 0529379 is underluminous (see below).

The Galactic extreme carbon stars do indeed extend the more local sample to longer periods 
and higher luminosities, larger expansion velocities and larger MLRs, and they better match the properties of the four LMC stars.
The gas-to-dust ratio needed to obtain the observed expansion velocity for the Galactic extreme C-stars shows two 
deviating values (67 and 477), but the five other stars show remarkably consistent values between 156 and 234, with a median of 191. 

For the LMC stars the situation is less obvious. Two stars require a value close to $\sim$135, while the other 
two need $r_{\rm gd} \sim 520$. 
Goldman et al. (2016) find a median value of 416 when using a similar technique to analyze OH/IR stars
A larger value might be expected as the metallicity is lower in the LMC, 
but C-stars dredge-up carbon from the interior in a way that is relatively insensitive of the initial metallicity and so the production of
C-rich dust may be larger with respect to the gas content in MC stars than in the Galaxy which would point to 
$r_{\rm gd}$ values lower than $\sim 200$.

The expansion velocities of the LMC C-stars are lower than most of the Galactic stars with comparable properties.
The range in Galactic C-star expansion velocities is larger than represented by the Danilovich et al. and Ramstedt \& Olofsson samples.
The sample of Groenewegen et al. (2002) contains expansion velocities for 309 C-stars, and 8 (or $\sim3$\%) have
velocities below 8.5 \ks, while 15\% have velocities above 25 \ks.
The median value of the expansion velocity of the stars with luminosities above 5400 \lsol\ (the lowest of the LMC stars) is 
17.6 \ks, while that for the 4 LMC C-stars is 11.4 \ks. 
Although the LMC sample is small, and the selection of, and the comparison to, a suitable galactic sample is non-trivial, 
it seems consistent with the finding for the Galactic halo C-stars that the expansion velocities are smaller at lower metallicity.
Hydrodynamical wind models by Wachter et al. (2008) support this trend.

The MLRs of the four stars turn out to be very similar and in agreement with the MLRs found for the Galactic extreme C-stars.
When plotted against luminosity there is one obvious outlier, ERO 0529379. Its luminosity is low in an absolute sense, but also
low for its pulsation period which is similar to IRAS 05125 and IRAS 05506 which do have similar high luminosities.
It is the only target with SiC in absorption (see Figure~\ref{Fig-SEDfits}), so its very large MLR is not in question.
One possibility is that the modelling assumption of spherical symmetry is not valid. 
This might be the case, but the width of the CO profile favours the assumption of expansion in a stellar outflow.
Keplerian discs are known in only two objects, the Red Rectangle (Bujarrabal et al. 2003) and 
AC Her (Bujarrabal et al. 2015), both post-AGB objects. Other objects may also have discs (Bujarrabal et al. 2013) 
but objects with discs are observationally characterised by narrow line profiles, 
with a width of \less~5 \ks\ in most cases (although theoretically the line profile may not necessarily 
be narrow as it depends on the inclination angle, see Homan et al. 2015). 

Sloan et al.\ (2016) showed that in the LMC, the pulsation amplitude of the carbon stars increases as they grow redder,
but it peaks before they have reached their reddest colors. They suggested that the stars beyond that peak may have 
evolved off of the AGB.  Two of the carbon stars in our sample are beyond that peak.  
Even if their long periods and the outflow velocity of ERO 0529379 imply that they could not have evolved 
far from the AGB, the low luminosity of ERO 0529379 suggests  that it may have a non-spherical geometry, 
which could be another indication of post-AGB evolution.

\subsection{ERO 0518117}

This source is peculiar in several ways. It was selected as being among the reddest sources known in the LMC (Gruendl et al. 2008).
It is one of five sources in the GS16 sample with an essentially featureless IRS spectrum, consistent with 
amorphous carbon dust without a trace of silicon carbide.
Such a featureless IRS spectrum is seen in R CrB stars (Kraemer et al. 2005) but the periodic light curve of ERO 0518117 
makes it unlikely that it is a R CrB star.

The CO observations now show that this object has the smallest expansion velocity of the 4 stars 
(but this may be statistically insignificant), and has a radial velocity of 164 \ks.
This is very atypical for the LMC for which Olsen et al. (2011) give a mean radial velocity of 263 \ks\ 
with a dispersion of 26 \ks, but coincides with the velocity of the SMC for which 
Dobbie et al. (2014) find a mean radial velocity of 150 \ks, and a dispersion of 26 \ks.
The star is located at about 4 degrees from the LMC center and at its location the rotation of the LMC disk 
is such (van der Marel et al. (2002), Olsen et al. 2011) that the line-of-sight velocity is about 220 \ks. 
This reduces the difference with the mean SMC velocity, but it is remains large.

It is unclear how to interpret these properties. Olsen et al. (2011) present evidence that the LMC has accreted stars from the SMC, but this
seems unlikely for this particular object. Evolved stars of this redness are not present among SMC stars in the GS16 sample, 
and specific searches for very dusty post-main-sequence objects in the MCs using far-infrared {\it Heritage} data 
(Meixner et al. 2013) have not turned up any such AGB stars in the SMC\footnote{For the object BMB-B 75 the far-IR 
emission is believed to be associated with a galaxy in the line-of-sight which would explain its 
redness in the mid-infrared.} (Jones et al. 2015).

\subsection{Evolutionary considerations}
\label{EvCo}

Ventura et al. (2016) investigated the nature of the most obscured C-rich AGB sources in the MCs. 
They combined the ATON and MONASH stellar evolution codes with a dust formation prescription, and they
concluded that the reddest C-rich sources have initial mass 2.5-3 \msol\ in the LMC 
(with $Z= 0.008$) and $\sim 1.5$ \msol\ in the SMC (with $Z= 0.004$). The difference between the galaxies is due to the difference 
in star formation histories. They predict that the LMC sources have [3.6$-$4.5] colours of 3.4, luminosities in the 
range 8~000 - 10~000~\lsol, MLRs of order 15 $\times 10^{-5}$~\msolyr, and have SiC in absorption.

These parameters are close to the two ERO sources which have [3.6$-$4.5] colours of 2.5-2.7, and ERO 0529379 which has SiC in absorption.
The difference in MLRs is a factor of 3, but this could be due to differences in the calculation of the absorption coefficients of the grains.

\begin{figure}
\centering

\includegraphics[width=0.99\hsize]{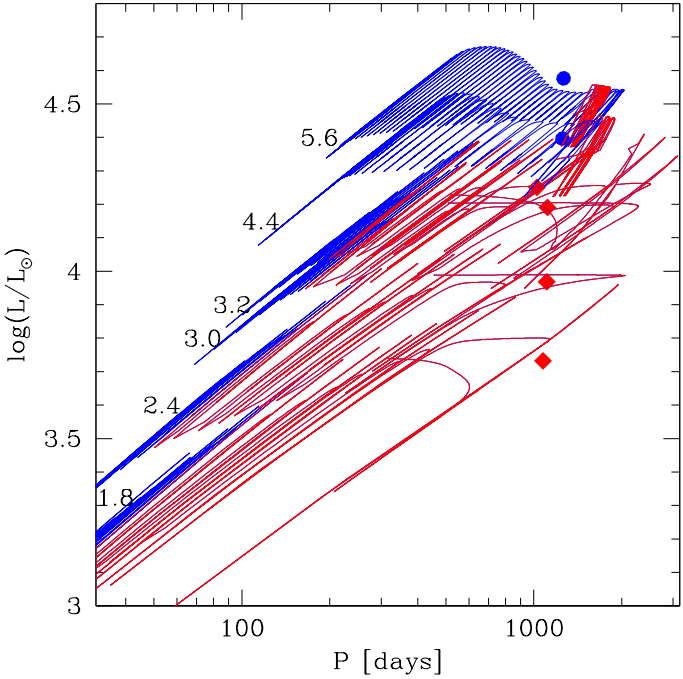}

\caption[]{ 
Evolution of the fundamental-mode period and luminosity during the TP-AGB based on the 
evolutionary models of Marigo et al. (2013) computed with the \texttt{COLIBRI} code. 
Initial masses (in $M_{\odot}$) are labelled, and the metallicity is $Z= 0.006$.
The stages with C/O$<$1 are marked in blue and with C/O$>$1 in red.
The observed points for the 2 OH/IR stars and 4 C-stars are marked by the blue circles, and the red diamonds, respectively.
} 

\label{Fig-Evolv} 
\end{figure}

Figure~\ref{Fig-Evolv} shows the evolution of the fundamental-mode period 
and luminosity during the TP-AGB phase for a few initial masses with 
metallicity $Z= 0.006$ based on the evolutionary models of Marigo et al. (2013) calculated with the \texttt{COLIBRI} code. 
Pulsation periods for the fundamental mode are calculated with the theoretical period-mass-radius relation given by equation~(5) of Wood (1990). 
Mass loss on the AGB is included  in \texttt{COLIBRI} using the formalism described by Marigo et al. (2013), 
updated as by Rosenfield et al. (2014, 2016), where the dust-driven regime follows an exponential increase 
as a function of stellar mass and radius. The results are similar to those obtained with the semi-empirical 
recipe introduced by Vassiliadis \& Wood (1993).

The third dredge-up is one of the most uncertain process in all TP-AGB models, so we opt to treat 
it with a parametric formalism.  Its efficiency is assumed following the results of 
full stellar AGB calculations of Karakas et al. (2002), with subsequent modifications 
as in Marigo \& Girardi (2007). In the models that experience hot-bottom burning (HBB), 
the third dredge-up is reduced compared to Karakas' original computations.

The changes in the surface chemical composition due to the third dredge-up and HBB are linked to the 
envelope and atmospheric structures through the on-the-fly computation of molecular chemistry and opacities 
with the \texttt{\AE SOPUS} code (Marigo \& Aringer 2009). This is important for a consistent 
prediction of the effective temperatures during the the TP-AGB phase.
The stages with C/O$<$1 are marked in blue and with C/O$>$1 in red.
The 1.8, 3.0 and 3.2 \msol\ models experience the third dredge-up, while the 4.4 and 5.6 \msol\ models 
experience the third dredge-up and HBB.
The observed points for the 2 OH/IR stars and the 4 C-stars are marked by the blue circles, and the red diamonds, respectively.

The TP-AGB models can explain the location of the observed stars in the P-L diagram.
The sizable excursions of the tracks both in luminosity and pulsation period are driven by the occurrence of
thermal pulses. In all models the evolution is followed until they reach the stage of lowest 
possible effective temperature (dictated by current mass and surface chemical composition), beyond which they
invert the trend moving towards higher temperatures and approaching the post-AGB stages. 
Regarding the C-stars, their long periods are predicted to be reached during the last thermal pulses when 
intense mass loss is stripping the envelope, the total mass is reduced (down to $1-2$~\msol), and the Hayashi 
evolutionary track for convective stars shifts towards lower effective temperatures, yielding larger radii hence longer periods.
The initial masses of the progenitors should be in the approximate range $2-3$~\msol, which agrees 
with Ventura et al. (2016), and, more generally, with the typical mass range for the formation of C stars
in the LMC, as indicated by their observed C-star luminosity function 
(Marigo \& Girardi et al. 2007). The present TP-AGB models predict the C-stars
will eventually become C-O white dwarfs with masses of $0.59-0.79$~\msol. 

The 2 bright OH/IR stars are interpreted as initially massive AGB stars in their final stages. 
The stars are well described by O-rich TP-AGB models with initial masses between $4.4-5.6$~\msol. 
For IRAS 05298 this is consistent with the initial mass estimate of $\sim$4~\msol\ based on its 
membership of a cluster (van Loon et al. 2001).
The OH/IR stars should have experienced HBB in the previous stages, passing through overluminous stages well 
above the core-mass luminosity relation, as shown by the typical bell-shaped luminosity tracks in Figure~\ref{Fig-Evolv}. 
We expect that HBB is currently extinguished after most of the envelope is ejected by stellar winds, 
and the actual masses are $\approx 2.0-2.6$~\msol. Models predict that the stars will 
soon (in a few 10~000 yr) become C-stars again due to a few final third dredge-up 
events (Frost et al. 1998, Marigo et al. 1998, van Loon et al. 1999),
and finally produce white dwarfs with masses of $0.95-1.05$~\msol. 
For these specific models the final C-star stages take place after the stars have reached 
the minimum effective temperature and start to warm towards the post-AGB phase.
The exact appearance of such late transition to the C-star domain critically depends 
on the efficiency of the third dredge-up. 

While an overall good agreement between models and observations on the P-L diagram is found, 
in a follow-up work we will perform a more detailed analysis to address various other aspects and their interplay 
(e.g. dust formation, radiative transfer across the dusty envelopes,
expansion velocities, efficiency of the third dredge-up, HBB, and mass loss).

\subsection{Gas MLRs}

An estimate of the gas MLR can be made using the formula in Ramstedt \& Olofsson (2008),
\begin{equation}
\dot{M} = s_{\rm J} \left( A \theta^2 D^2 \right)^{a_{\rm J}} {\rm v}_{\rm exp}^{b_{\rm J}} \, f_{\rm CO}^{c_{\rm J}}
\label{Eq-CO}
\end{equation}
with, for the J= 2-1 transition, $s_{\rm J}= (1.3 \pm 0.7) \cdot 10^{-11}$~\msolyr, $a_{\rm J}= 0.82$, $b_{\rm J}= 0.46$, $c_{\rm J} = 0.59$.
The distance, $D$, is set to 50~kpc, and the averaged beam size, $\theta$, is 1.43\arcsec. The integrated intensity, $A$, 
is taken from Table~\ref{Tab-Fit}.

The only parameter to fix is $f_{\rm CO}$, the CO abundance relative to H$_2$.
The \texttt{COLIBRI} code introduced in the previous section also predicts the abundances of various elements as a funtion of eveolutionary timescale.
For a star of initial mass 2.4 \msol, what we believe is the mass of the extreme red C-stars, $f_{\rm CO}$ is roughly 
constant at $4.5 \cdot 10^{-4}$ from the thermal pulse where the star turns into a C star until the end of the AGB evolution.
This value basically depends only on the assumed intial oxygen abundance in the models.
For massive intermediate-mass stars the time evolution of the CO abundance is much more complex due to HBB (which converts carbon into nitrogen), 
which is a function of time on the AGB (traced in the models by the pulsation period as an observable proxy of time), 
details of the modelling of HBB, and initial mass.
For a star of 4.4 \msol, typical of the mass of the 2 OH/IR stars, at a pulsation period of about 1300 days, $f_{\rm CO}$ is 
about $2.0 \cdot 10^{-4}$, but is uncertain by at least a factor of 2. 
The CO-based MLRs are listed in the last column of Tab.~\ref{Tab-Fit}. 
These MLRs are compared to the dust-based MLRs in the last column of Table~\ref{TabDUSTY} for the C-stars, and to the 
dust-based MLRs from Goldman et al. (2016) for IRAS 05298 ($8.4 \cdot 10^{-5}$ \msolyr) and IRAS 04545 ($7.1 \cdot 10^{-5}$ \msolyr).

These values systematically fall below those based on consistent modelling of the dust. 
For the OH/IR stars the discrepancy is a factor $>$~20, 
for the C-stars it is a factor of $4-5$ (with the exception of IRAS 05506 where the 
CO-based MLR is larger than that based on the dust modelling).
In deriving Eq.~\ref{Eq-CO}, Ramstedt \& Olofsson did assume typical values of galactic sources for the free parameter 
$h = \left(\frac{100}{r_{\rm gd}}\right) \left(\frac{2}{{\rho}_{\rm d}}\right) 
\left(\frac{0.05 \mu m}{a_{\rm d}}\right)$, with $a_{\rm d}$ the dust grain radius, in their models.
A systematic difference of this parameter between the solar neighbourhood and LMC will likely have an effect on this relation.

If Eq.~\ref{Eq-CO} is applicable and the dust-based MLRs are correct then the CO abundances must be 
substantially lower (by an order of magntiude) than assumed, and/or the observed CO intensities 
are lower by a factor of $\sim$7 for the C-stars and $>30$ for the OH/IR stars than expected based on Eq.~\ref{Eq-CO}.
The dust emission and the CO J= 2-1 emission trace different radii in the envelope, so one possible explanation for this 
difference between dust and CO based mass-loss rate is that it varies in time, a suggestion first made by 
Heske et al. (1990) in connection with Galactic OH/IR stars.
Another possibility is that the interstellar radiation field (ISRF) is stronger in the LMC. 
Paradis et al. (2009) mention that, assuming the same spectral shape for the ISRF as in the solar neighborhood, 
the ISRF in the diffuse LMC medium is $\sim$~5 times stronger. The strength of the [C{\sc ii}] line relative 
to CO in LMC star-forming regions is much larger than in Milky Way ones, also pointing 
to a stronger radiation field (Israel \& Maloney 2009).
Recent work by McDonald et al. (2015) and Zhukovska et al. (2015) have demonstrated the importance of the strength 
of the ISRF on the expected CO emission in clusters. That the largest difference between CO-based and dust-based MLR 
is observed for the IRAS 05298, which is known to be in a cluster, is consistent with this. 
The influence of the ISRF and a full line radiative transfer modeling is deferred until the CO J= 3-2 lines 
have been measured by ALMA.

\section{Summary and conclusions}

The first results are presented of a programme that that aims to accurately determine gas and dust MLRs in AGB stars in the MCs.
The keys to success are detection and modelling of CO thermal emission lines.
In this paper we present observations of the CO J= 2-1 line of 2 OH/IR and 4 carbon stars in the LMC.
The OH/IR stars are weaker than anticipated in this line, and only one OH/IR is marginally detected. 

The detection of all 4 C-stars in the CO J= 2-1 line allowed us to determine the expansion velocities and compare them to a
sample of Galactic C-stars, including ``extreme'' C-stars that have similar large MLRs as the LMC targets.
Determining the gas MLR directly and through RT modelling is deferred until the CO J= 3-2 data 
are in hand, and in this paper the MLR is determined from fitting the SED and {\it Spitzer}/IRS spectrum 
using the DUSTY code in the mode where the hydrodynamical equations 
of dust and gas are solved and the expansion velocity and gas MLR are predicted.

These first results support the conclusions previously derived from CO observations of 
metal-poor C-stars in the Galactic Halo: at lower metallicity the expansion velocity in C stars is smaller, 
and the MLR similar, to a comparable star at solar metallicity. 
It must be stressed, however, that our sample is small and finding a suitable sample of comparison stars 
in the Galaxy is challenging. Therefore, this conclusion awaits testing with improved samples 
both within the Galaxy and in the Magellanic Clouds.

\acknowledgements{  
This paper makes use of the following ALMA data:
ADS/JAO.ALMA\#2013.1.00319.S. ALMA is a partnership of ESO
(representing its member states), NSF (USA) and NINS (Japan), together
with NRC (Canada) and NSC and ASIAA (Taiwan) and KASI (Republic of Korea), 
in cooperation with the Republic of Chile. 
The Joint ALMA Observatory is operated by ESO, AUI/NRAO and NAOJ.
MATG would like to thank the Nordic ALMA regional centre for their hospitality and support in the data reduction and analysis.
JBS wishes to acknowledge the support of a Career Integration Grant 
within the 7th European Community Framework Program, FP7-PEOPLE-2013-CIG-630861-SYNISM. 
LD acknowledges support from the ERC consolidator grant 646758 AEROSOL and the FWO Research Project grant G024112N.
MWF gratefully acknowledges the receipt of research grants from the National Research Foundation of South Africa (NRF).
OCJ acknowledges support from NASA grant, NNX14AN06G. 
PM acknowledges support from the ERC Consolidator Grant funding scheme (project STARKEY, G.A. n.~615604).
MM is supported by the STFC Ernest Rutherford fellowship.
HO acknowledges financial support from the Swedish Research Council.
RS's contribution to the research described here was carried out at the 
Jet Propulsion Laboratory (JPL), California Institute of Technology, under a contract with NASA.
GCS was supported by the NSF through Award 1108645.
JvL acknowledges support from the UK Science and Technology Facility Council under grant ST/M001040/1.
WV acknowledges support from the ERC through consolidator grant 614264.
AZ and IM acknowledge supported from the UK Science and Technology Facility Council under grant ST/L000768/1.
}

{}

\begin{appendix}

\section{DUSTY fits to the LMC C-stars}
\label{AppDUSTY}

\begin{figure*}
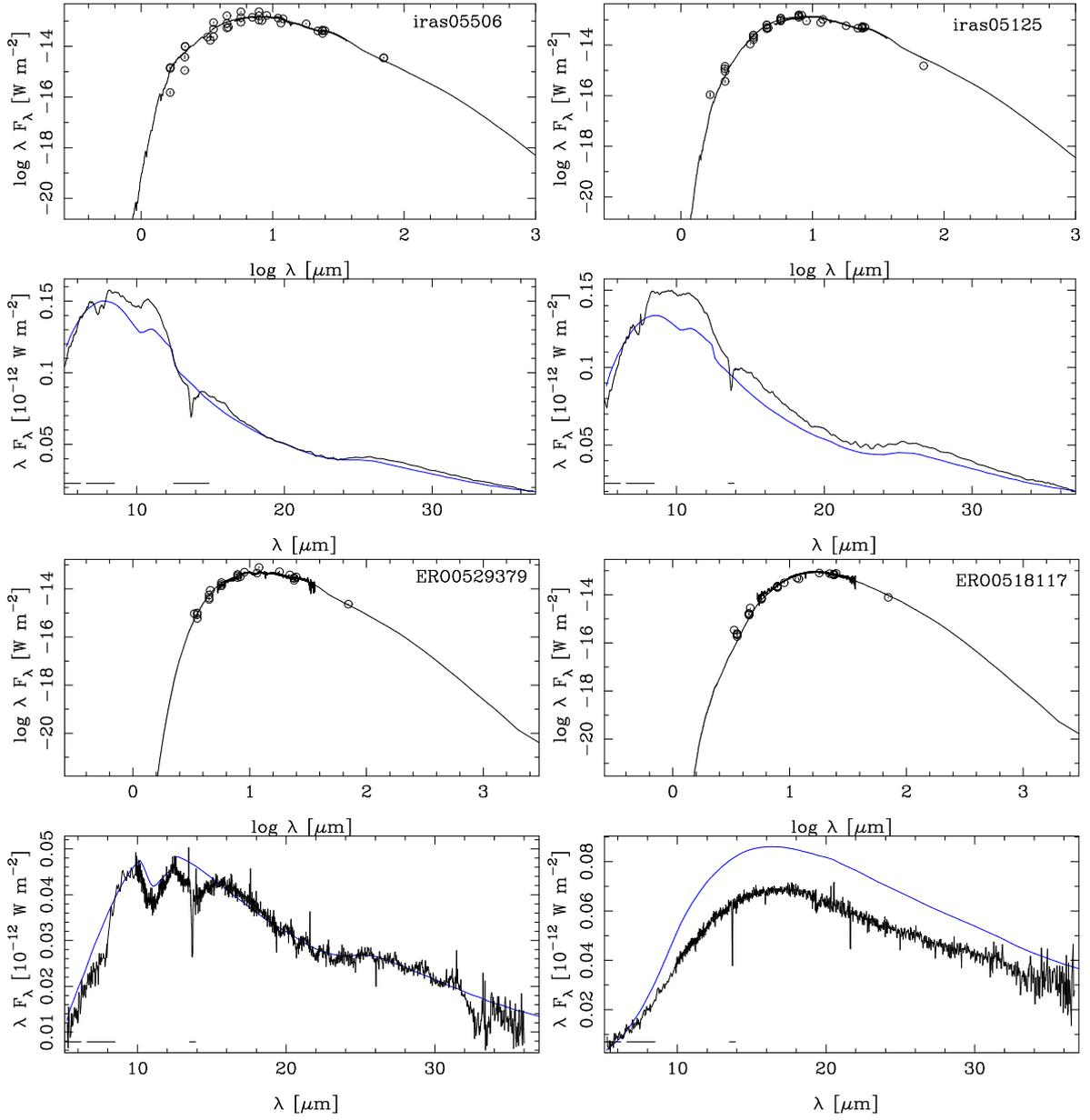

\centering

\begin{minipage}{0.42\textwidth}
\resizebox{\hsize}{!}{\includegraphics[angle=-0]{iras05506_sed.ps.20161761122}} 
\end{minipage}
\begin{minipage}{0.42\textwidth}
\resizebox{\hsize}{!}{\includegraphics[angle=-0]{iras05125_sed.ps.20161761123}} 
\end{minipage}

\begin{minipage}{0.42\textwidth}
\resizebox{\hsize}{!}{\includegraphics[angle=-0]{ERO0529379_sed.ps.20161760954}} 
\end{minipage}
\begin{minipage}{0.42\textwidth}
\resizebox{\hsize}{!}{\includegraphics[angle=-0]{ERO0518117_sed.ps.20161760955}} 
\end{minipage}

\label{Fig-SEDfits}

\caption{Fits to the SED and IRS spectrum of the C-stars. 
The small vertical lines near the bottom of the plot with the spectrum indicate wavelength regions excluded from the fitting. 
}
\end{figure*}

Figure~\ref{Fig-SEDfits} shows the fits to the SED and IRS spectra of the 4 C-stars when running DUSTY in ``density mode = 3'' mode.
The top panel shows the photometry and IRS spectrum in a $\log - \log$ plot together with the model.
The bottom panel shows the IRS spectrum and the model (the blue line). 
The model is scaled to the observation based on the average flux in the $6.4-6.6$~$\mu$m region.
The small vertical lines near the bottom of the plot indicate wavelength regions excluded from the fitting.

\newpage

\section{DUSTY fits to the Galactic extreme C-stars}
\label{AppDUSTYExt}

Figure~\ref{Fig-SEDfitsGal} shows the fits to the SED and {\it IRAS}/LRS or {\it ISO}/SWS spectra of the seven 
Galactic extreme C-stars when running DUSTY in ``density type = 3'' mode.

\begin{figure}[H]
\centering

\includegraphics[width=0.8\hsize]{afgl190_sed.ps.20162081026}

\caption{Fits to the SED and mid-IR spectrum of Galactic ``extreme'' C-stars. }
\label{Fig-SEDfitsGal}
\end{figure}

\setcounter{figure}{0}
\begin{figure*}
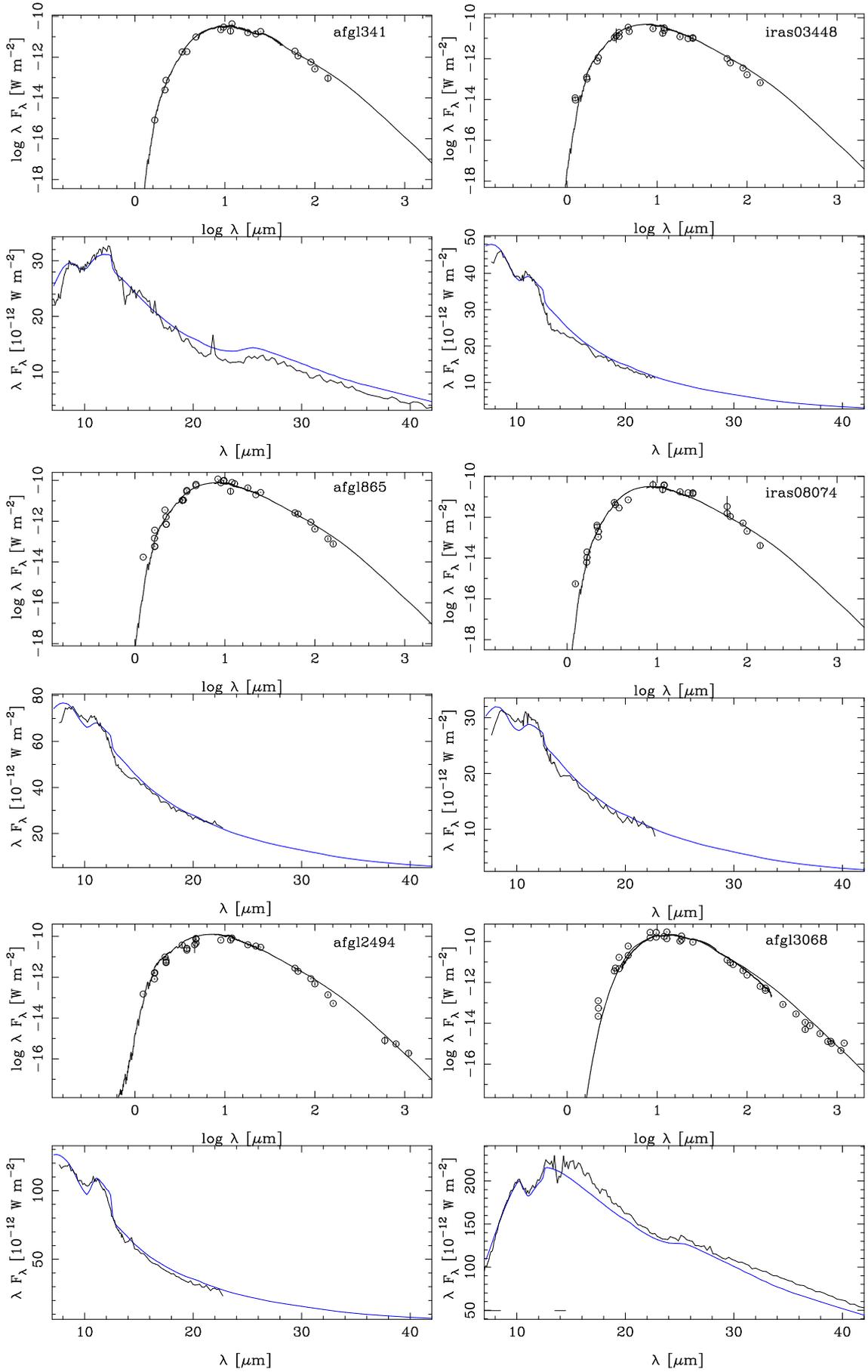

\centering

\begin{minipage}{0.405\textwidth}
\resizebox{\hsize}{!}{\includegraphics[angle=-0]{afgl341_sed.ps.20162081027}} 
\end{minipage}
\begin{minipage}{0.405\textwidth}
\resizebox{\hsize}{!}{\includegraphics[angle=-0]{iras03448_sed.ps.20162071457}} 
\end{minipage}

\begin{minipage}{0.405\textwidth}
\resizebox{\hsize}{!}{\includegraphics[angle=-0]{afgl865_sed.ps.20162081027}} 
\end{minipage}
\begin{minipage}{0.405\textwidth}
\resizebox{\hsize}{!}{\includegraphics[angle=-0]{iras08074_sed.ps.20162081026}} 
\end{minipage}

\begin{minipage}{0.405\textwidth}
\resizebox{\hsize}{!}{\includegraphics[angle=-0]{afgl2494_sed.ps.20162081027}} 
\end{minipage}
\begin{minipage}{0.405\textwidth}
\resizebox{\hsize}{!}{\includegraphics[angle=-0]{afgl3068_sed.ps.20162081028}} 
\end{minipage}

\caption{Fits to the SED and mid-IR spectrum of Galactic ``extreme'' C-stars, continued.}
\end{figure*}

\end{appendix}

\end{document}